\documentclass{article}
\usepackage{graphicx}
\usepackage{amsmath}
\usepackage{algorithm}
\usepackage{algpseudocode}
\usepackage{textcomp}
\usepackage{xcolor}
\usepackage{acronym}
\usepackage{todonotes}
\usepackage{booktabs}
\usepackage{hyperref}
\usepackage{caption}
\usepackage{subcaption}
\usepackage{amssymb}
\usepackage[perpage]{footmisc}
\usepackage{arxiv}

\title{Transmission Topology Optimization using accelerated MapElites}

\author{ \href{https://orcid.org/0009-0004-1761-4305}{\includegraphics[scale=0.06]{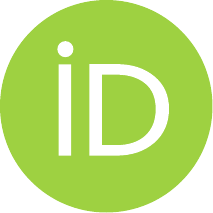}\hspace{1mm}Nico Westerbeck} \\
	Elia Group\\
    Brussels\\
    Belgium\\
	\And
	\href{https://orcid.org/0009-0006-4756-2889}{\includegraphics[scale=0.06]{orcid.pdf}\hspace{1mm}Leonard Hilfrich} \\
	Elia Group\\
    Brussels\\
    Belgium\\
    \And
	\href{https://orcid.org/0000-0002-3623-5341}{\includegraphics[scale=0.06]{orcid.pdf}\hspace{1mm}Dirk Witthaut} \\
	Forschungszentrum Jülich\\
    Jülich\\
    Germany\\
}

\acrodef{QD}[QD]{Quality-Diversity}
\acrodef{TTO}[TTO]{Transmission Topology Optimization}
\acrodef{OTS}[OTS]{Optimal Transmission Switching}
\acrodef{MILP}[MILP]{Mixed-Integer Linear Programming}
\acrodef{RL}[RL]{Reinforcement Learning}
\acrodef{DACF}[DACF]{Day-Ahead Congestion Forecast}
\acrodef{IDCF}[IDCF]{Intra-Day Congestion Forecast}
\acrodef{SC-ACOPF}[SC-ACOPF]{Security Constrained AC Optimal Power Flow}
\acrodef{TSO}[TSO]{Transmission System Operator}
\acrodef{IGM}[IGM]{Individual Grid Model}
\acrodef{SSA}[SSA]{Static Security Analysis}
\acrodef{PST}[PST]{Phase Shifting Transformer}
\acrodef{HVDC}[HVDC]{High-Voltage Direct Current transmission}
\acrodef{PTDF}[PTDF]{Power Transfer Distribution Factors}

\begin{document}
\maketitle
\begin{abstract}

Transmission Topology Optimization has great potential to improve efficiency and flexibility of grid operations through non-costly switching actions, but previous approaches struggle with runtime performance and scalability. In this work, we present an optimization approach that leverages GPU acceleration to speed up computations. In a genetic algorithm setting, topologies are randomly mutated and evaluated in parallel for multiple optimization criteria. Combined with a fully GPU-native DC loadflow solver, there is no CPU-GPU data transfer required in the DC optimization loop. Using a variant of the illumination algorithm MapElites, we efficiently generate a set of diverse candidate solutions on the pareto front. Together with an importing and AC validation step, we present an end-to-end optimization solution that runs in under 15 minutes. The approach is currently under evaluation by operational planning operators in two European TSOs. We furthermore open-source our code at github.com/eliagroup/ToOp.

\end{abstract}
\section{Introduction}

% Background: Congestion due to renewables
The energy transition challenges power system operation in multiple ways. In particular, the increasing integration of renewable energy sources can increase transmission grid loads, as generation is often located in remote areas with favorable conditions~\cite{titz2026cascading}. Furthermore, the growing dependence on weather introduces significant temporal variability in grid utilization~\cite{staffell2018increasing}.

% Focus on Germany & costs
These challenges are especially pronounced in Germany, where large shares of wind generation are installed in the North and offshore, while major load centers are located in the West and South of the country~\cite{titz2023identifying}. This spatial mismatch results in frequent congestion and requires costly interventions by transmission system operators. German congestion management costs amounted to €3.3 billion in 2023 and €2.9 billion in 2024~\cite{siemonsmeier2019congestion,bundesnetzagentur2026monitoringbericht}. In addition, redispatch is often provided by conventional power plants~\cite{titz2023identifying}, leading to increased greenhouse gas emissions.

% Countermeasures
Grid expansion is planned to mitigate congestion in the long term, but implementation is slow. \acf{TTO} offers a complementary, cost-effective approach~\cite{granelli2006optimal,hedman2011review}. By reconfiguring switches, power flows can be redirected away from congested lines toward less utilized parts of the network. While this typically does not eliminate congestion entirely, it can significantly reduce the need for conventional congestion management measures such as redispatch and countertrading.

% The challenges
However, the practical implementation of \ac{TTO} is challenging. First, finding an optimal or close-to-optimal grid configuration in the large combinatorial search space represents a huge computational challenge. More precisely, it has been shown that \ac{TTO} is a NP-hard problem~\cite{kocuk2017new}. Second, grid reconfiguration has to respect a multitude of operational constraints~\cite{westerbeck2026contraints}, which are difficult to incorporate into mathematical programming.

% This paper
In this work, we propose a GPU-based approach to \ac{TTO} using a MapElites algorithm inspired by QDax~\cite{chalumeau2023qdax,mouret2015illuminating} combined with an accelerated DC power flow solver from our previous work~\cite{westerbeck2025accelerated}. The optimization is implemented fully on GPU with minimal communication overhead, enabling the evaluation of tens of thousands of candidate topologies per second. The approach is embedded in a three-stage workflow, including a preprocessing step to reduce problem complexity and an AC validation stage to ensure feasibility. Overall, the method allows to load, optimize, and validate congested grid states in under 15 minutes, making it suitable for operational use. The approach is currently being tested by operational planners at two European \acp{TSO}. Our code is openly available at \url{github.com/eliagroup/ToOp}.

\section{Literature}

% Big picture: High potential but difficult
Several studies demonstrate the significant potential of \ac{TTO} for reducing congestion and operational costs in power systems with a high penetration of renewable generation~\cite{little2021optimal,numan2023role,lichtinghagen2026innoptem,EPRI_TransmissionTopologyOptimization_2024}. However, large-scale deployment in real-world transmission systems remains limited. The primary challenge lies in the combinatorial complexity of the problem: for realistic grids with thousands of nodes and tens of thousands of switching elements, the number of feasible topologies grows exponentially, rendering exhaustive or naive search approaches intractable. As a result, the design of scalable and efficient optimization algorithms is a central research challenge in \ac{TTO}.

% Computational complexity, MILP and RL
The problem of \ac{TTO}, often called \ac{OTS} \cite{hedman2009optimal, hedman2011review} is a combinatorial optimization problem that attempts to find the optimal switching state of a power grid with respect to an optimization criterion. Initially, disconnecting branches was a prime focus, with later works incorporating bus splitting and reassignments \cite{zhou2021substationlevel,babaeinejadsarookolaee2023transmission}. The \ac{TTO} problem is often combined with redispatch optimization to form a unified congestion management optimizer \cite{ewerszumrode2024iterative,lichtinghagen2026innoptem}. \ac{MILP} approaches tend to struggle with high runtime especially on large grids \cite{lehmann2014complexity,pineda2024tight}, and multiple ideas are presented how to reduce runtime \cite{ruiz2012tractable,li2023solving,niu2014review}. Alternative ideas utilize \ac{RL} to learn a policy in an expensive ahead-of-time training step, which can be used for rapid inference in grid situations that match the training data distribution \cite{marchesini2025rl2grid,sar2025optimizing,subramanian2021exploring,lautenbacher2025multiobjective}. Also, genetic algorithm approaches have been submitted but traditionally struggle to evaluate sufficiently many combinations to reach good performance levels \cite{granelli2006optimal}.

% From genetic algorithms to MapElites
Genetic algorithms \cite{holland1984genetic} are a form of gradient-free optimization algorithms. They successively apply random changes to a repertoire of candidate solutions, evolving them towards a fitness maximum. Such gradient-free approaches generally require large numbers of evaluations and perform worse in domains where good gradients are readily available. \acf{QD} algorithms \cite{lehman2011evolving} apply the concept to multi-objective optimization, aiming to illuminate the pareto-front efficiently instead of finding a single maximum point. Approaches like NSGA \cite{deb2002fast,deb2014evolutionary} or Deep MapElites \cite{flageat2020fast} extend the search beyond just the first pareto front. \ac{QD} algorithms tend to be parallelized easily \cite{lim2022accelerated} and multiple frameworks offer GPU implementations \cite{chalumeau2023qdax,lange2022evosax}. 

In this work, we present a three-stage GPU accelerated approach at topology optimization. Combining a GPU native DC loadflow solver \cite{westerbeck2025accelerated} and a GPU native MapElites implementation inspired by QDax \cite{chalumeau2023qdax, mouret2015illuminating}, we implement a topology optimizer that resides fully on GPU and requires minimal PCI communication during the optimization loop. The resulting performance makes it viable to evaluate tens of thousands of topology candidates per second. We pair that with a preprocessing stage that simplifies the problem without much loss of generality and an AC validation stage which ensures viability of the solutions. Together, we present a high performance \ac{TTO} solution that is able to load, optimize and validate a congested grid in under 15 minutes. Our solution is currently being tested in operational planning at two european \acp{TSO}. Our code is openly available at \url{github.com/eliagroup/ToOp}.

\section{Background}

\subsection{Market design and congestion management}

In the European context, the practical deployment of \ac{TTO} must be understood within the framework of zonal electricity markets. The European power system operates under a zonal market design, where bidding zones are coupled through the Single Day-Ahead Coupling (SDAC) mechanism~\cite{trebbien2024patterns}. While cross-zonal transmission constraints are explicitly considered during day-ahead market clearing, intra-zonal constraints are neglected. As a consequence, significant congestion can arise within bidding zones, which must be resolved ex post by \acp{TSO} through redispatch and countertrading measures~\cite{bertsch2017relevance,hirth2018marketbased}. This issue is particularly pronounced in Germany and Luxembourg, which form a single bidding zone despite substantial internal transmission bottlenecks~\cite{titz2023identifying}, but is a pressing issue for most \acp{TSO} operating in zonal markets.

To ensure secure system operation, \acp{TSO} rely on established congestion management procedures in an operational planning framework such as the \ac{DACF} or \ac{IDCF}. These follow a sequential workflow, where congestion is first identified and then resolved through corrective actions such as topology, redispatch and countertrading. Topology as a corrective measure is particularly interesting, as switching actions are non-costly; both redispatch and countertrading incur significant cost penalties. Currently, both TSOs in this study employ a manual topology optimization stage, where operators identify viable topologies based on experience and expert assessment. A \ac{TTO} solution in the form of a recommender system must be carefully integrated so as not to interfere with existing process timings. In particular, topology measures are required to be determined prior to the major stages of redispatch planning, leaving very little time for an optimization to complete.

\subsection{Operational constraints for topology optimization}
\label{sec:operational}

In current operational procedures, topology proposals are considered based on manual assessment and expert knowledge and validated through AC power flow studies. An automated \ac{TTO} system is expected to substantially extend the decision space, finding safe and effective remedial actions. The validation of these actions should happen regarding current, voltage, short-circuit, dynamic stability and other operational constraints \cite{westerbeck2026contraints}. However, performing all these validations is computationally prohibitive and instead we only validate \ac{SSA} information. For the current scope of work, we rely on the system operators to perform additional studies if necessary.

Operational practice imposes strict time constraints on any machine-assisted \ac{TTO} approach. In Germany, the manual topology optimization within the model refinement phase typically takes on the order of 45 minutes. In Belgium, operators strive to come up with topology solutions in under 30 minutes. Any potential topology adjustment must be selected and evaluated during this period with enough time for potential additional studies. A \ac{TTO} system with a human in the loop setup must leave enough time for the human to review proposals - especially if dynamic stability and short-circuit computations are skipped as in our approach. In practice, this implies that optimization results must be available within at most 15 minutes, with shorter runtimes being highly desirable.

These operational constraints have important implications for the design of \ac{TTO} methods. In particular, they favor fast, robust, and interpretable approaches that can be integrated into existing sequential workflows. In line with operational practice, non-costly optimization can be performed prior to the optimization of other remedial actions. This motivates a two-stage approach, in which a topology optimization proposes candidate configurations that are subsequently evaluated and complemented by other measures such as redispatch. In this work, we focus on designing such an approach that can be directly integrated into existing planning procedures.

Treating topology optimization as a stand-alone step changes the nature of the problem formulation. Constraints that would appear as hard constraints in a fully integrated optimization — such as branch flow or voltage magnitude limits — can instead be handled as soft criteria at this stage. This is because final feasibility is ensured in a subsequent step through redispatch and countertrading. Consequently, the objective of topology optimization is not to fully resolve all violations, but to identify grid configurations that reduce the severity and cost of corrective actions in the following stages.

\subsection{Switching actions in topology optimization}

In this work, we focus on a subset of switching-based actions:
\begin{enumerate}
    \item Branch disconnections: A line or transformer that feeds power into a congestion is disconnected.
    \item Busbar reassignments: Assets are assigned to another busbar prior to a busbar coupler opening.
    \item Busbar coupler openings (\texttt{splits}): The coupler(s) connecting two or more busbars is opened, electrically separating them.
\end{enumerate}
These actions are jointly referred to as \texttt{de-meshing} actions, as they reduce connectivity in the grid and redirect power flows toward alternative paths. Branch disconnections remove a line or transformer that contributes to a congestion, thereby redistributing power flows across the remaining network. Busbar reassignments modify the connection of assets to busbars within a substation, typically via disconnector switching. Busbar coupler openings (\texttt{splits}) electrically separate busbars by opening the connecting load-breaker. As busbar reassignments only affect power flows when combined with a busbar split, we treat both actions jointly as \texttt{reconfigurations}. In this work, we restrict ourselves to the common case of two-node operation and do not consider splits into more than two electrically separate components.

Additional non-costly measures such as the cancellation of planned maintenances (\texttt{reconnections}), \ac{PST} or \ac{HVDC} setpoint adjustments are also used in operational practice together with de-meshing actions. Integrating these into a \ac{TTO} framework to form a general non-costly optimizer is sensible, we leave this for future work to tackle.

Performing a large number of switching actions can significantly reduce redispatch requirements, but also moves the grid away from the well-understood fully meshed state into less secure de-meshed states. 

A trade-off between aggressive and conservative switching and associated risk profiles must be found. As it is difficult to assign explicit costs to all relevant dimensions of this trade-off, we instead aim to present the Pareto front across select objective dimensions within predefined feasibility boundaries. We focus on different aspects of switching distance as our descriptor dimensions.

\section{Approach}

We present an accelerated \ac{QD} approach to efficiently tackle the transmission topology optimization problem as defined above. The proposed solution is structured into three components: an \textbf{Importer}, a \textbf{DC-Optimizer}, and an \textbf{AC-Validator}.

The Importer performs preprocessing and simplifies the problem as much as possible ahead of time. Most steps of the importing process can be executed before the exact forecasts are available, relying on only the initial grid topology. In particular, an action set is constructed, from which candidate switching actions are sampled during optimization.

The DC-Optimizer conducts a large-scale topology search based on the DC power flow approximation. Its purpose is to efficiently explore the combinatorial search space and identify promising topology combinations from all candidates in the action set. Leveraging the accelerated DC solver presented in~\cite{westerbeck2025accelerated}, the optimizer can evaluate a large number of configurations within a short time.

The AC-Validator does not generate new topologies but evaluates candidate solutions proposed by the DC-Optimizer using a full AC power flow model. It acts as a quality gate before results are presented to operators. As the number of promising DC candidates exceeds the number of AC evaluations feasible within the available runtime, a dominance-based selection strategy is applied to identify a diverse and representative subset of solutions for validation.

\subsection{Problem formulation and Notation}

We consider a power grid represented by a graph with node set $\mathcal{N}$ and edge set $\mathcal{E}$, with cardinalities $|\mathcal{N}| = N_n$ and $|\mathcal{E}| = N_e$. The system is subject to a set of injections $\mathcal{I}$, with $|\mathcal{I}| = N_i$, representing generation and load connected to the nodes.

Security is ensured in a \ac{SSA} by evaluating the grid under a set of outage cases $\mathcal{O}$, with $|\mathcal{O}| = N_o$. These include line and generator contingencies as well as simple multi-element outages such as three-winding transformers or pylon failures. Busbar outages are treated separately and denoted by a set $\mathcal{O}_b$, with $|\mathcal{O}_b| = N_{ob}$. For simplicity, we refer to the grid under any such contingency as an \texttt{N-1 state}, even though some contingencies such as busbar outages may remove multiple elements.

Let $\mathcal{N}_s \subseteq \mathcal{N}$ denote the set of switchable nodes, with $|\mathcal{N}_s| = N_{ns}$. For these nodes, detailed node-breaker information is available, allowing the identification of physically feasible reconfigurations. Typically, nodes within the core region of a \ac{TSO} are fully switchable, while nodes in the surrounding observation area are not.

Each edge $e \in \mathcal{E}$ is subject to a flow limit $\bar{f}_e$. Furthermore, only a subset of edges $\mathcal{D} \subseteq \mathcal{E}$ is eligible for disconnection. Together with the reconfigurations, these elements define the feasible action space for topology optimization. 

\subsection{Fitness function heuristics}

The fundamental aim of \ac{TTO} is to reduce the costs of congestion management. As discussed in Sec.~\ref{sec:operational}, operational practice follows a two-stage approach in which topology optimization precedes the planning of redispatch and countertrading. Consequently, topology optimization requires a surrogate measure of congestion that serves as a proxy for redispatch effort.

We therefore define a vector of heuristic metric scores $\lambda$ that can be computed directly from power flow results and capture different aspects of congestion and solution quality. These metrics are independent of the underlying power flow model (AC or DC).
We denote:
\begin{itemize}
    \item $f^{n0} \in \mathbb{R}^{N_e}$: branch flows in the base case (N-0),
    \item $f^{(o)} \in \mathbb{R}^{N_e}$: branch flows under contingency $o \in \mathcal{O}$,
    \item $f^{(b)} \in \mathbb{R}^{N_e}$: branch flows under busbar outage $b \in \mathcal{O}_b$,
    \item $\bar{f} \in \mathbb{R}_+^{N_e}$: admissible branch flow limits.
\end{itemize}
To assess N-1 security, we define for each edge $e \in \mathcal{E}$ the maximum post-contingency loading:
\[
f_e^{\max} = \max_{o \in \mathcal{O}} |f_e^{(o)}|,
\]
and collect these values in the vector $f^{\max}$.

The N-1 security criterion requires $f_e^{\max} \leq \bar{f}_e$ for all $e \in \mathcal{E}$. To quantify congestion, we measure the extent to which these conditions are violated using the clip operator $\Pi(x) = \max(x, 0)$.
We thus define the following congestion metrics:
\begin{itemize}
    \item N-1 absolute overload energy:
    \[
    \lambda_o = \sum_{e \in \mathcal{E}} \Pi(f_e^{\max} - \bar{f}_e)
    \]
    
    \item N-1 number of critical branches:
    \[
    \lambda_c = \sum_{e \in \mathcal{E}} \mathbb{1}(f_e^{\max} > \bar{f}_e)
    \]
    
    \item N-0 number of critical branches:
    \[
    \lambda_{c0} = \sum_{e \in \mathcal{E}} \mathbb{1}(f_e^{n0} > \bar{f}_e)
    \]
\end{itemize}

Busbar outages are treated separately. Analogous to the N-1 overload metric, we define a busbar penalty:
\[
\lambda_b = \sum_{e \in \mathcal{E}} \Pi\left(\max_{b \in \mathcal{O}_b} |f_e^{(b)}| - \bar{f}_e \right).
\]
This metric is compared to the corresponding value $\lambda_b^{pre}$ of the pre-optimization topology and used as a do-not-make-it-worse constraint in the optimization.

In addition to flow-based metrics, we define measures that quantify the deviation from the reference fully meshed topology and the effort required to reach a given configuration. These measures can be directly computed from any given topology without needing to solve the power flows:
\begin{itemize}
    \item $\lambda_d$: number of branch disconnections,
    \item $\lambda_s$: number of substations with splits,
    \item $\lambda_r$: number of reassignments required to reach the topology.
\end{itemize}

\subsection{Importer}

The importer is one of the three major components. It performs preprocessing steps to convert grid models from the CGMES format into an internal representation and enumerate the action set. CGMES (Common Grid Model Exchange Standard) is a data format defined by ENTSO-E for the exchange of power system models across European transmission system operators~\cite{gietz2019cgmes}. 

As part of the preprocessing, a linearized DC representation of the grid is constructed and the corresponding \ac{PTDF} matrix is computed for the pre-optimization topology. This allows rapid evaluation of branch flows for different injection patterns and topologies, which is exploited by the DC-optimizer as described in the following section.

A central task of the importer is the construction of the action space for topology optimization. We distinguish between two classes of actions: branch disconnections and substation-local reconfigurations. The set of disconnectable branches is denoted by $\mathcal{D}$, and the set of substation-local reconfigurations (including coupler openings) by $\mathcal{A}$. The set $\mathcal{D}$ contains all branches whose removal does not create islands in the grid under the base case (N-0) or any contingency $o \in \mathcal{O}$. This is checked using Tarjan's bridge-finding algorithm \cite{tarjan1974note}.

Entries in $\mathcal{A}$ correspond to station-local combinations of busbar reassignments and coupler openings. Each action includes at least one opened busbar coupler and, potentially, additional asset reassignments within the same station. While, in principle, all combinations of switch states could be considered, the size of $\mathcal{A}$ grows exponentially with the number of assets. To keep the problem tractable, the following simplifications are applied:
\begin{enumerate}
    \item The importer processes each substation independently, generating and storing only station-local reconfigurations; global combinations across substations are constructed at runtime.
    \item Only electrically distinct configurations are retained, i.e., configurations that result in different bus/branch representations of the grid.
    \item Only splits resulting in two-node operation are included.
    \item Configurations that create islands under the base case (N-0) or any contingency $o \in \mathcal{O}$ are excluded. This is checked using Tarjan's bridge-finding algorithm \cite{tarjan1974note}.
    \item Configurations that violate asset connectivity constraints are excluded.
    \item Configurations that are not reachable in the node-breaker model are excluded.
    \item If a station has more than $2^{23}$ feasible configurations, a random subset is retained.
\end{enumerate}

The enumeration of $\mathcal{A}$ is performed in two stages. First, all electrically distinct bus/branch representations of a station that result in two-node operation are generated. This corresponds to up to $2^{N_{e,s} + N_{i,s} - 1}$ combinations, where $N_{e,s}$ and $N_{i,s}$ denote the number of branches and injections at the station, respectively. The limit of $2^{23}$ configurations per station is enforced at this stage.

In a second step, each bus/branch representation is mapped to a feasible node-breaker realization using heuristic procedures. Configurations for which no valid realization can be found are discarded. For the remaining actions, the reassignment distance $\lambda_r$ in the node-breaker model is precomputed and stored in a lookup table.

\subsection{DC-Optimizer}

The main task of the DC optimizer is to scan the large search space for promising candidates and forward them to the AC validator. Solving the full non-linear AC power flow equations is orders of magnitude slower, rendering a direct search on the AC model infeasible. However, recent advances in accelerated DC loadflow computations \cite{westerbeck2025accelerated, vandijk2024bus, vandijk2024unified} enable the evaluation of large numbers of candidate topologies under a linearized approximation.

The employed DC solver builds on a precomputed \ac{PTDF} matrix obtained in the importer stage. Using this representation, branch flows can be computed as linear functions of nodal injections, while topology changes such as line outages or substation reconfigurations are incorporated via low-rank updates to the \ac{PTDF} matrix \cite{vandijk2024unified}. These operations can be executed efficiently in parallel on GPUs, enabling very high throughput. As a result, large batches of loadflow evaluations can be performed efficiently, making a broad exploration of the topology space feasible~\cite{westerbeck2025accelerated}.

To exploit this capability, we design an accelerated MapElites algorithm \cite{mouret2015illuminating} inspired by the QDax framework \cite{chalumeau2023qdax}. MapElites belongs to the class of quality-diversity algorithms and extends classical genetic algorithms by not only searching for a single optimal solution, but by maintaining a diverse set of high-quality solutions across a predefined descriptor space. The entire evolutionary process is implemented on the GPU, avoiding costly host-device communication during optimization iterations. Only promising candidate topologies are transferred from device to host memory for subsequent AC validation.

\subsubsection{Repertoire design}

\begin{figure}
    \centering
    \includegraphics[width=0.8\linewidth]{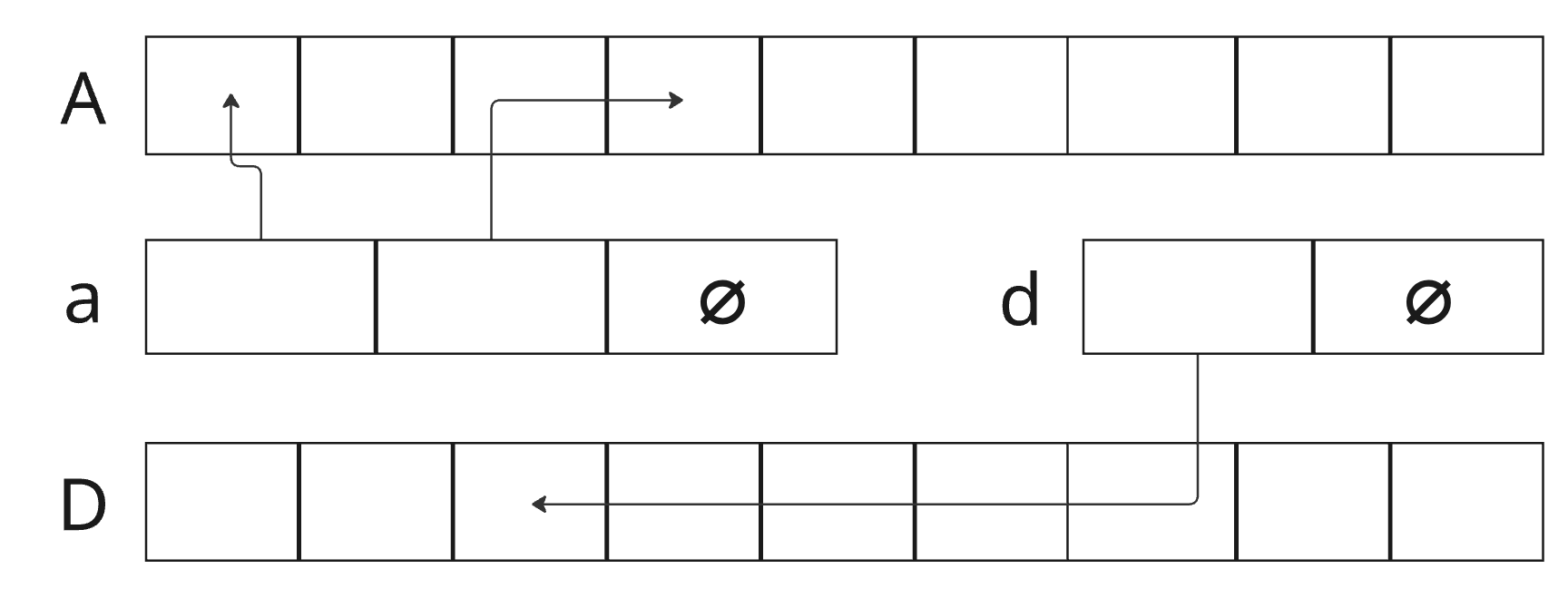}
    \caption{An illustration of the genome. Two out of three action slots are set to action 1 and 4 in the action set $\mathcal{A}$, one out of two disconnection slots is set to entry 3 in the disconnection set $\mathcal{D}$. The other slots are filled with the empty action $\varnothing$.}
    \label{fig:genome}
\end{figure}

In quality-diversity optimization, a repertoire denotes a collection of candidate solutions that jointly represent diverse and high-performing behaviors. In the context of \ac{TTO}, each candidate corresponds to a specific combination of topological switching actions and line disconnections.

To be more precise, the repertoire $\mathbf{R} = (a, d)^{N_r}$ contains $N_r$ individuals, where each individual represents a combination of topological actions on the grid. Formally, an individual is defined as a tuple $(a^{N_a} \in A \cup \varnothing, d^{N_d} \in D \cup \varnothing)$, where $N_a$ denotes the number of action slots and $N_d$ the number of disconnection slots. All $a$ are selected from distinct substations, and all $d$ target distinct branches. Empty split and disconnection slots are represented by $\varnothing$. The setup is illustrated in Figure~\ref{fig:genome} for $|A| = 9$, $|D| = 9$, $N_a = 3$, and $N_d = 2$.

The MapElites algorithm introduces descriptor dimensions into the repertoire. Descriptors separate the repertoire into cells, where individuals are competing only in their respective cell. Loosely speaking, each cell evolves independently toward a different region of the solution space, thereby promoting diversity among candidate solutions. In particular, this allows for individuals with non-optimal overall fitness to remain in the repertoire if no other individual with the same descriptors can outperform them -- effectively illuminating the Pareto front.

In this work, we define the repertoire cells using the topology metrics $\lambda_d$, $\lambda_s$, and $\lambda_r$. These descriptors characterize the switching distance with respect to the default grid configuration and thereby encode different levels of operational intervention. As a result, the repertoire contains solutions ranging from conservative topologies close to the default configuration to more strongly reconfigured network states.

Rather than prescribing an explicit trade-off between topology risk and optimization objective, the MapElites formulation preserves high-performing individuals across all descriptor combinations. This provides system operators with a diverse set of candidate solutions that differ not only in fitness, but also in their degree of deviation from standard operation.

For the practical implementation, the repertoire cells are labeled as $0, 1, \ldots, N_r-1$. We define a mapping function $\beta(\lambda) \rightarrow 0, \ldots, N_r-1$ that assigns each individual to its corresponding cell according to the scoring vector $\lambda$. In the following, we assume $\lambda_{d} \in \{0,1,2\}$, $\lambda_{r} \in \{0,\ldots,45\}$, and $\lambda_{s} \in \{0,1,2,3\}$, resulting in a repertoire size of $N_r = 3 \times 46 \times 4 = 552$ cells. One possible mapping is given by
\[
\beta(\lambda) = \lambda_{d} + 3(\lambda_{s} + 46 \lambda_{r}) .
\]

\subsubsection{Mutation Operation}

\begin{algorithm}
\caption{Mutation}
\label{alg:mutation}
\begin{algorithmic}[1]
\State For a given topology $(a, d)$

\Function{Add}{$x, X, \phi$}
  \State Sample $i \sim \mathrm{Uniform}(|x|)$ s.t. $x_i = \varnothing$ and $* \sim \mathrm{Uniform}(X \setminus \phi)$
  \State Set $x_i \gets *$, return $x$
\EndFunction

\Function{Remove}{$x, X$}
  \State Sample $i \sim \mathrm{Uniform}(|x|)$ s.t. $x_i \neq \varnothing$
  \State Set $x_i \gets \varnothing$, return $x$
\EndFunction

\Function{Change}{$x, X, \phi$}
    \State Sample $* \sim \mathrm{Uniform}(X \setminus \phi)$
    \State If action mutation then: $i$ s.t. $x_i \in s(*)$ else: $i \sim \mathrm{Uniform}(|x|)$ s.t. $x_i \neq \varnothing$
  \State Set $x_i \gets *$, return $x$
\EndFunction

\Function{Identity}{$x$}
  \Return x
\EndFunction

\Comment{Substation mutation}
\State Sample $N_{\mathrm{mut}} \sim \Pi(\mathrm{Poisson}(N_{\mathrm{mutmean}}), 1, N_a)$

\For{$r = 1$ to $N_{\mathrm{mut}}$}
    \State Sample a feasible operation:
    \State \quad $\circ \sim \mathrm{Weighted}(\{\textsc{Add}, \textsc{Remove}, \textsc{Change}, \textsc{Identity}\})$
    \State Parametrize $\circ: $
    \State \quad $\textsc{Add}(a, A, s(a))$
    \State \quad $\textsc{Remove}(a, A$) 
    \State \quad $\textsc{Change}(a, s(a), a)$
    \State \quad $\textsc{Identity}(a)$
    \State Apply $a' \gets \circ$
\EndFor

\Comment{Disconnection mutation}

\State Sample a feasible operation
\State \quad $\circ \sim \mathrm{Weighted}(\{\textsc{Add}, \textsc{Remove}, \textsc{Change}, \textsc{Identity}\})$
\State If $(a, d) = (\varnothing^{N_a}, \varnothing^{N_d})$ then set $\circ \gets \textsc{Add}$
\State Parametrize $\circ:$
\State \quad $\textsc{Add}(d, D, d)$
\State \quad $\textsc{Remove}(d, D)$
\State \quad $\textsc{Change}(d, D, d)$
\State \quad $\textsc{Identity}(d)$
\State Apply $d' \gets \circ$

\State \Return $(a', d')$

\end{algorithmic}
\end{algorithm}

% General concept
Mutation is the main exploration mechanism in MapElites and evolutionary optimization. Starting from an existing individual, mutation generates new candidate solutions by applying random modifications. In the context of \ac{TTO}, this corresponds to changing switching actions and line disconnections of a topology candidate. The overall mutation procedure is summarized in Alg.~\ref{alg:mutation}.

% Constraints and notation
Not all modifications result in feasible topologies. In particular, a substation must not be split multiple times and a branch cannot be disconnected more than once. To express these constraints, we define the function $s(a)$, which returns all actions affecting the same substation as action $a$. Furthermore, we define an exclusion set $\phi$ that prevents invalid samples during mutation. For disconnection sampling, the exclusion set is given by $\phi=d$, ensuring that no branch is disconnected twice. For action sampling, we use $\phi=s(a)$ to prevent multiple splits of the same substation.

% Four operations
Both action and disconnection mutation are based on four elementary operations: \textsc{Add}, \textsc{Remove}, \textsc{Change}, and \textsc{Identity}. \textsc{Add} replaces an empty slot with a newly sampled action. \textsc{Remove} clears an existing slot. \textsc{Change} replaces an existing entry with a different feasible action, while \textsc{Identity} leaves the slot unchanged. Newly sampled actions and disconnections must satisfy the feasibility constraints described above.

% Repeated mutation
Action mutation is repeated $N_{\mathrm{mut}} \in \{1,\ldots,N_a\}$ times, allowing multiple substations to change during a single mutation step. Instead of using a fixed value, $N_{\mathrm{mut}}$ is sampled from a Poisson distribution with mean $N_{\mathrm{mutmean}} = 2$. After the action mutation stage, one disconnection mutation is applied.

% Sampling probabilities
In every mutation step, one feasible operation is randomly selected among \textsc{Add}, \textsc{Remove}, \textsc{Change}, and \textsc{Identity} according to predefined probabilities. For action mutation, we use $p_{add}=0.2$, $p_{change}=0.5$, $p_{remove}=0.2$, and $p_{identity}=0.1$. For disconnection mutation, we use $p_{add}=0.25$, $p_{change}=0.5$, $p_{remove}=0.25$, and $p_{identity}=0$. If an individual contains neither a split nor a disconnection, a disconnection is forced to avoid reproducing the unchanged default topology.

\subsubsection{Crossover Operation}

\begin{algorithm}
\caption{Crossover}\label{alg:crossover}
\begin{algorithmic}[1]
\State Given two individuals $a^1, d^1$ and $a^2, d^2$
\State Initialize $a' \gets \varnothing^{N_a}$, $d' \gets \varnothing^{N_d}$
\State
\Comment{Action crossover}
\For{$i$ in $N_a$}
    \State Sample $a'_i \sim \mathrm{Weighted}((a^1 \cup a^2) \setminus a')$ s.t. $p(a'_i \in a^1) = p_{c1}$
\EndFor
\State
\Comment{Disconnection crossover}
\For{$i$ in $N_d$}
    \State Sample $d'_i \sim \mathrm{Weighted}((d^1 \cup d^2) \setminus d')$ s.t. $p(d'_i \in d^1) = p_{c1}$
\EndFor

\end{algorithmic}
\end{algorithm}

Crossover is a second exploration mechanism commonly used in evolutionary optimization. While mutation modifies a single individual, crossover combines information from two existing individuals to generate a new candidate solution. In the context of \ac{TTO}, this corresponds to combining switching actions and line disconnections from two parent topologies.

The crossover procedure is summarized in Alg.~\ref{alg:crossover}. Similar to mutation, crossover consists of an action and a disconnection component. Starting from two parent individuals $(a^1,d^1)$ and $(a^2,d^2)$, the algorithm iteratively samples feasible actions and disconnections from the union of both parents until the offspring individual is constructed.

Crossover introduces a bias parameter $p_{c1}$ that controls the preference toward the first parent individual. For $p_{c1}=1$, crossover reproduces $(a^1,d^1)$ exactly, whereas $p_{c1}=0.5$ corresponds to unbiased sampling from both parents. Intermediate values allow the offspring to inherit features preferentially from one parent while still introducing diversity through recombination.

\subsubsection{High-level Genetic Algorithm loop}

\begin{algorithm}
\caption{Accelerated MapElites for TTO}\label{alg:highlevel}

\begin{algorithmic}[1]
\State initialize $\mathbf{R} \gets \varnothing^{N_a},\varnothing^{N_d}$
\While{runtime $<$ limit}
   \For{iter per epoch}
    \State sample mutation-crossover ratio $b_{mcratio} \gets 0..b$
    \State perform Alg. \ref{alg:mutation} SIMD parallel with $b_{mcratio}$ individuals sampled from $R$
    \State perform Alg. \ref{alg:crossover} SIMD parallel with $b - b_{mcratio}$ pairs of individuals sampled from $R$
    \State
    \State Compute N-1 DC loadflows SIMD parallel for new individuals
    \State Compute $\lambda$ and $\beta$ SIMD parallel
    \State
    \Comment{unbatched}
    \State sorted insert into the cell in $\mathbf{R}$ according to $\beta$
    \EndFor
    \State On CPU: send $\mathbf{R}$ to AC optimizer via message queue
\EndWhile
\end{algorithmic}
\end{algorithm}

The overall optimization workflow is summarized in Alg.~\ref{alg:highlevel}. Starting from an initial repertoire, the algorithm repeatedly generates, evaluates, and reinserts new topology candidates.

Every iteration consists of four stages. First, new candidate topologies are generated through mutation and crossover as described above. Second, DC load flow and contingency analyses are computed for all newly generated individuals using the methods described~\cite{westerbeck2025accelerated}. Third, the resulting topology metrics are evaluated and aggregated into a scalar fitness value depending on the study case
\[
\begin{aligned}
\lambda_{f1} &= -\left(\lambda_{o} + 200\,\lambda_{c0} + 50\,\lambda_{c}\right), \\
\lambda_{f2} &= -\left(\lambda_{o} + 200\,\lambda_{c0} + 50\,\lambda_{c}
+ \Pi(\lambda_{b} - \lambda_{b}^{\mathrm{pre}})\right).
\end{aligned}
\]

which is maximized by the optimization. With the given formulation, the fitness will always be negative, however we stay consistent with conventions in the field of \ac{QD} by \textit{maximizing fitness} instead of \textit{minimizing penalty}. The score primarily targets the reduction of overload energy $\lambda_o$, while additionally penalizing critical branches under N-0 and N-1 conditions through $\lambda_{c0}$ and $\lambda_c$. As $\lambda_o$ and $\lambda_b$ are measured in MW and range in the thousands in our use-cases, they dominate the reward function even if the critical branch count (usually <10) is added with a factor of $50$ or $200$ respectively. Furthermore, the final term discourages deteriorating busbar outage behavior compared to the pre-optimization state. Finally, the evaluated individuals are inserted back into the repertoire according to their descriptor mapping $\beta$, where they compete only against individuals within the same cell.

To fully exploit GPU acceleration, all computationally intensive steps are executed in SIMD-parallel batches. In particular, mutation, crossover, and DC load flow evaluation operate simultaneously on batches of candidate topologies. To maintain constant tensor shapes during GPU execution, all operations are performed on padded inputs of size $b$. Only the repertoire insertion step remains unbatched. Independent random seeds are used for each batch element to ensure stochastic diversity, although duplicate topology evaluations may still occur with low probability.

In order to minimize GPU-CPU communication overhead, multiple iterations are grouped in an \textit{epoch}. Only after the end of each epoch, results are sent back to the AC optimizer via a message queue. Sending the results to the AC optimizer after every genetic algorithm iteration would result in large overhead, as the GPU execution loop needs to be interrupted for a GPU-CPU transfer of the results every time. In this work, we group 500 iterations into one epoch, resulting in topology results being sent every few seconds.

\subsection{AC-Validator}

The DC optimizer yields a diverse set of candidate topologies that potentially reduce grid congestion. Operational security guidelines require that the transmission grid remains N-1 secure at all times. Since the DC power flow model used during optimization is only an approximation of the physical system, it cannot guarantee feasibility or security with sufficient accuracy. Therefore, every candidate topology must undergo a full AC \ac{SSA} before it can be presented to operators for final assessment.

Performing AC N-1 security analysis for large transmission grids is computationally expensive, making it infeasible to validate all candidate topologies generated by the DC optimizer within the available runtime. To address this challenge, we introduce three performance optimizations: CPU parallelization, a worst-$k$ contingency case selector, and a repertoire elimination process.

We utilize the security analysis feature in the open-source grid modelling framework powsybl \cite{lf-energypowsybl} for all AC loadflow computations. This framework natively comes with CPU parallelization for the loadflow kernel. Furthermore, we utilize polars CPU-parallelized LazyFrame feature \cite{polars} to detect constraint violations in the large amount of loadflow data returned by the loadflow kernel.

The idea behind the worst-k contingency cases filter is to use information from the DC N-1 analysis to guide the selection of N-1 cases in AC. Usually, only a few contingency cases dominate the result. For this, the DC Optimizer computes for every N-1 case individually the overload energy and writes out the worst $k$ N-1 cases. We set $k=20$ in this work. The AC validator then first computes these N-1 cases only and discards the topology early if it has worse overloads or poor convergence. If it can not be discarded, a full N-1 analysis follows. Unfortunately, the CPU parallelization feature of powsybl is ineffective for $k=20$ cases and we thus disable it on the worst-k filter.

Furthermore, a repertoire elimination process is employed which selects DC candidates for validation. The elimination utilizes the following three heuristics to prune the number of candidates for evaluation.
\begin{enumerate}
    \item Don't AC validate topologies where there is a very similar topology that was validated already
    \item Don't AC validate if there is a dominating topology with lower switching distance but similar or better DC fitness.
    \item Don't AC validate if the fitness improvement in DC is less than a pre-configured threshold.
\end{enumerate}
If no more candidates remain after pruning, the validation randomly chooses remaining topologies. 

Using CPU parallelization, worst contingency case filtering and repertoire elimination, our AC validation stage evaluates DC-proposed topologies effectively.

\section{Results}

In this section, we evaluate the proposed optimization framework on European transmission grids using real-world operational data. To facilitate this, we introduce two study grids derived from operational planning scenarios and then investigate the performance of the proposed approach from three perspectives.

First, we analyze the allocation of computational budget with particular emphasis on achieving end-to-end runtimes below 15 minutes. Next, we investigate repertoire evolution over time. Finally, we assess optimization quality by comparing congestion-related metrics, in particular overload energy, before and after optimization.

\begin{table}[ht]
\centering
\caption{Benchmark Grids}
\label{tab:grids}
\begin{tabular}{lccccc}
\toprule
  & $N_e$ & $N_o$ & $N_{ns}$ & $|A|$ & $|D|$ \\
\midrule
TSO 1 & 1526 & 752 & 67 & 168.596 & 318 \\ 
TSO 2 & 872 & 365 & 50 & 19.070 & 145 \\ 
\bottomrule
\end{tabular}
\end{table}

\subsection{Benchmark Grids}

We evaluate the proposed approach on two European transmission grids using data from the operational planning processes of the respective \ac{TSO}. Both scenarios contain congestion situations without pre-configured topological remedial actions. Topology optimization shall initially mitigate the occurring overloads before any remaining congestion is handled through redispatch processes.

Each grid model consists of the detailed \ac{IGM} of the respective \ac{TSO}, including all substations where switching actions may be applied. TSO 1 holds as well a reduced representation of bordering transmission and distribution systems. The resulting grid sizes are summarized in Table~\ref{tab:grids}.

Despite these reduced study cases, the combinatorial complexity of the optimization problem remains extremely large. For example, considering only configurations with up to $N_a=3$ switching actions and $N_d=2$ disconnections for \ac{TSO}~1, and disregarding the constraint that actions must originate from distinct substations, the search space already contains
\[
\binom{168\,596}{3} \times \binom{318}{2} \approx 10^{19}
\]
possible topology candidates. For the TSO 1 grid, we maximize $\lambda_{f1}$ and for the TSO 2 grid $\lambda_{f2}$. The TSO 2 grid operators require a do-not-make-it-worse type of busbar contingency analysis, while the TSO 1 grid is currently evaluated without busbar contingencies. However, we want to note that defining the fitness function $\lambda_f$ is part of the operational evaluation process and likely subject to change.

At this point we want to emphasize that these study grids are concerned each with a single timestep of a historic day in the \ac{DACF}, and the same optimization performed on different timesteps of the same grid yields vastly different results.

\subsection{Performance and efficiency}

To assist the operators, the results must be available in a timely manner, as the day-ahead process has a tight time window for topology optimization. To integrate seamlessly with operational processes, a \ac{TTO} system must run in less than 15 minutes end-to-end. This means that the limited compute budget needs to be spent carefully on those computations that matter. To assess this, we evaluate exploration coverage and the runtime performance of different components of the application.

\begin{table}[t]
\centering
\caption{Average number of splits and disconnections in a batch with 64 topologies, on the TSO 2 study case.}
\label{tab:split_disc_count}
\begin{tabular}{c|ccc}
 & \multicolumn{3}{l}{\textbf{Disconnections}} \\
\textbf{Splits} & $\lambda_d = 0$ & $\lambda_d = 1$ & $\lambda_d = 2$ \\
\cline{2-4}
$\lambda_s = 0$ & 0 & 2 & 1 \\
$\lambda_s = 1$ & 5 & 5 & 5 \\
$\lambda_s = 2$ & 7 & 9 & 8 \\
$\lambda_s = 3$ & 7 & 8 & 7 \\
\end{tabular}
\end{table}

Given the size of the search space, the computational budget needs to be allocated with care. Ideally, the algorithm should achieve a broad and balanced coverage of the Pareto front while distributing computational resources effectively across the repertoire cells. Table \ref{tab:split_disc_count} shows the average number of individuals per batch on the TSO 2 study case at batch size 64, separated by number of splits and disconnections. We can observe that all possible combinations are explored, with a bias for higher amount of splits and disconnections. Since the search space increases exponentially with more splits and disconnections, it is desirable to allocate more compute budget to these regions. The low number of individuals with only disconnections and $\lambda_s = 0$ is similarly desirable, as the computational complexity of just the disconnection action space is lower than that of the reconfigurations.

\begin{figure*}
    \centering
    \includegraphics[width=1.\linewidth]{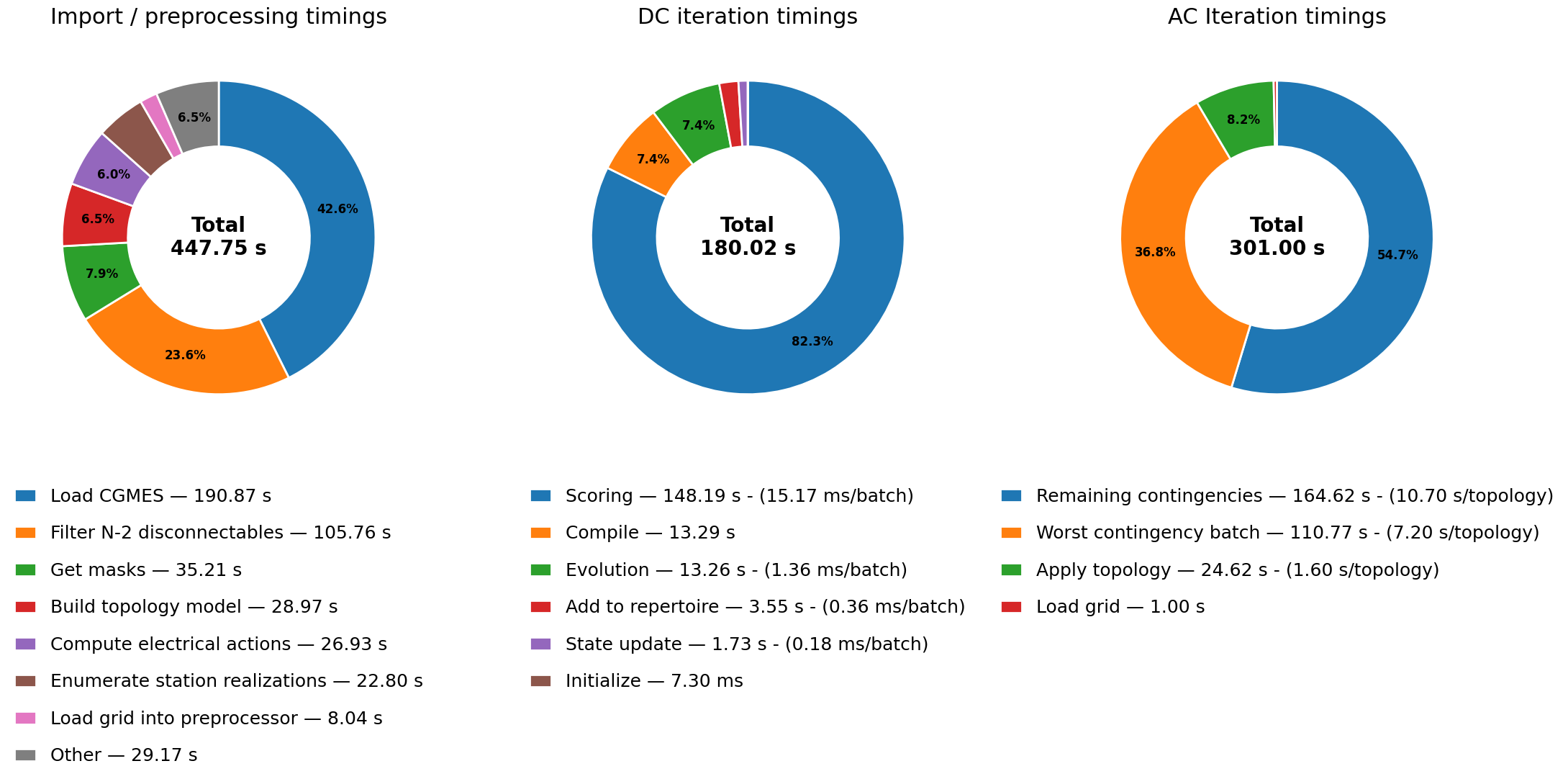}
    \caption{Runtime performance of different components of the application for a representative optimization setting on the TSO 1 grid.}
    \label{fig:profiling}
\end{figure*}

In Figure \ref{fig:profiling} we investigate the runtime performance of the three components. We plot the total runtime of each component along with the percentage of the underlying methods. On the importer side, parsing the CGMES file takes up a large percentage of the total runtime. This is unique to the data situation at hand where the grid needs to be recombined from more than 50 individual profiles. With a total runtime of 7 minutes, the import lowers the available runtime for the DC and AC stages down to 8 minutes to hit the 15 minute target. An additional 2 minutes are lost on optimizer startup and visualization (not shown here). Hence, we set the AC runtime to 5 minutes and because the last DC topologies have a low chance to still be validated by AC we stop the DC solver after 3 minutes already so it is free for the next optimization.

The DC optimizer spends most of its time scoring topologies, which is done through loadflow computations by the accelerated DC loadflow solver presented in \cite{westerbeck2025accelerated}. The evolution algorithm presented in this work, consisting of the core evolution, repertoire addition and state update only takes up ca 10\% of the total optimizer runtime. While loading the data is quick, the DC optimizer uses jax which JIT-compiles the code for 13.29 seconds in the first optimization epoch. Choosing a precompiled language like CUDA would not incur this penalty, but comes with higher development complexity.

The AC solver spends 36.8\% computing only the worst-k contingencies and 54.7\% for full N-1 analysis runs. Even though the worst k run only has to compute 20 contingencies while the full N-1 run has 789 contingencies, they have almost the same runtime per topology. This hints at further improvement potential in the AC validation pipeline. The small number of contingencies in the worst-k variant causes the CPU parallelization feature of the framework to be ineffective and some runtime is lost in constant-time data copy operations. 

% Question 3: Are the results good?
\subsection{Result Quality}

For measuring the result quality, we count topologies that underwent a full N-1 security analysis and passed all acceptance criteria. As the DC optimization stage runs on multiple heuristic assumptions, it is to be expected that a part of the topologies do not pass on AC. Table \ref{tab:ac_accepted_count} shows how many topologies have passed the AC validation pipeline, separated by how many disconnections $\lambda_{d}$ and how many splits $\lambda_s$ are contained in the topology. First, it can be noticed that a higher number of complex topologies (large $\lambda_s$, $\lambda_d$) pass the validation stage. This can partially be explained by the observation from Table \ref{tab:split_disc_count} that fewer simple topologies are sampled by the DC mutation. But also there seems to be more flexibility in the solution space, especially with 1 or 2 splits and at least 1 disconnection.

\begin{table}[t]
\centering
\caption{Number of AC-accepted topologies for different cells in the repertoire on the TSO 2 study scenario. The pre-optimization topology at (0, 0) never entered the AC validation and is not counted here.}
\label{tab:ac_accepted_count}
\begin{tabular}{c|ccc}
 & \multicolumn{3}{c}{\textbf{Disconnections}} \\
\textbf{Splits} & $\lambda_d = 0$ & $\lambda_d = 1$ & $\lambda_d = 2$ \\
\cline{2-4}
$\lambda_s = 0$ & 0 & 1 & 2 \\
$\lambda_s = 1$ & 2 & 22 & 21 \\
$\lambda_s = 2$ & 3 & 21 & 25 \\
$\lambda_s = 3$ & 10 & 11 & 15 \\
\end{tabular}
\end{table}

\begin{table}[t]
\centering
\caption{AC N-1 overload energy $\lambda_o$ in [MW] of the best topology for every given number of splits and disconnections, on both study cases. The pre-optimization overload is added at (0, 0) for both grids.}
\label{tab:overload_heatmap}
\begin{tabular}{c|ccc}
 & \multicolumn{3}{c}{\textbf{Disconnections}} \\
\textbf{Splits} & $\lambda_d = 0$ & $\lambda_d = 1$ & $\lambda_d = 2$ \\
\cline{2-4}
\multicolumn{4}{c}{\emph{TSO 1}} \\
\cline{2-4}
$\lambda_s = 0$ & 1269 & - & - \\
$\lambda_s = 1$ & - & - & 160 \\
$\lambda_s = 2$ & 163 & 160 & 160 \\
$\lambda_s = 3$ & 160 & 160 & 160 \\
\cline{2-4}
\multicolumn{4}{c}{\emph{TSO 2}} \\
\cline{2-4}
$\lambda_s = 0$ & 2637 & 1416 & 918 \\
$\lambda_s = 1$ & 2171 & 1351 & 909 \\
$\lambda_s = 2$ & 2182 & 1417 & 905 \\
$\lambda_s = 3$ & 1730 & 1457 & 908 \\
\end{tabular}
\end{table}

Furthermore, we investigate whether this flexibility also reflects in better solution quality. In Table \ref{tab:overload_heatmap}, we measure the primary metric performance $\lambda_o$ of the different AC topologies for the two study grids. 

On the TSO 1 grid in Table \ref{tab:overload_heatmap}, a pre-optimization overload energy of 1269MW is shown at $\lambda_s=0$, $\lambda_d=0$. We encounter a scenario where no topologies with just one split or disconnection are found that resolve the problem. The simplest topology that lowers the overloads has at least two splits, or at least one split and two disconnections. All solutions with only one split or one disconnection are rejected. This connects with one of the fundamental problems of topology optimization, where topology actions can be non-additive. Only the combination of two actions benefits the grid situation but each individual action makes it worse. The decision whether to use the split-based or disconnection-based solutions with similar result quality is dependent on the operational situation in the control room. Feasibility of implementation might vary between the topologies, and a manual analysis taking into account the grid situation before and after the congestion, operator workload and other constraints is currently the best way to choose a final candidate.

On the TSO 2 grid in Table \ref{tab:overload_heatmap}, we encounter a situation where disconnections are very effective while bus splits are not. The cell with 0 splits and 0 disconnections represents the pre-optimization performance at an overload energy of 2637MW. The best found topology has 905MW and uses two disconnections and two bus splits. However, another topology is found that reaches 918MW and avoids bus splits entirely. The difference of 13MW is small and barely exceeds typical forecast errors. In operations, the topology with slightly higher overload energy would likely be preferred due to its simplicity. Every additional switching action introduces operational risk and must be well justified.

\begin{figure}
\centering
\includegraphics[width=1.\linewidth]{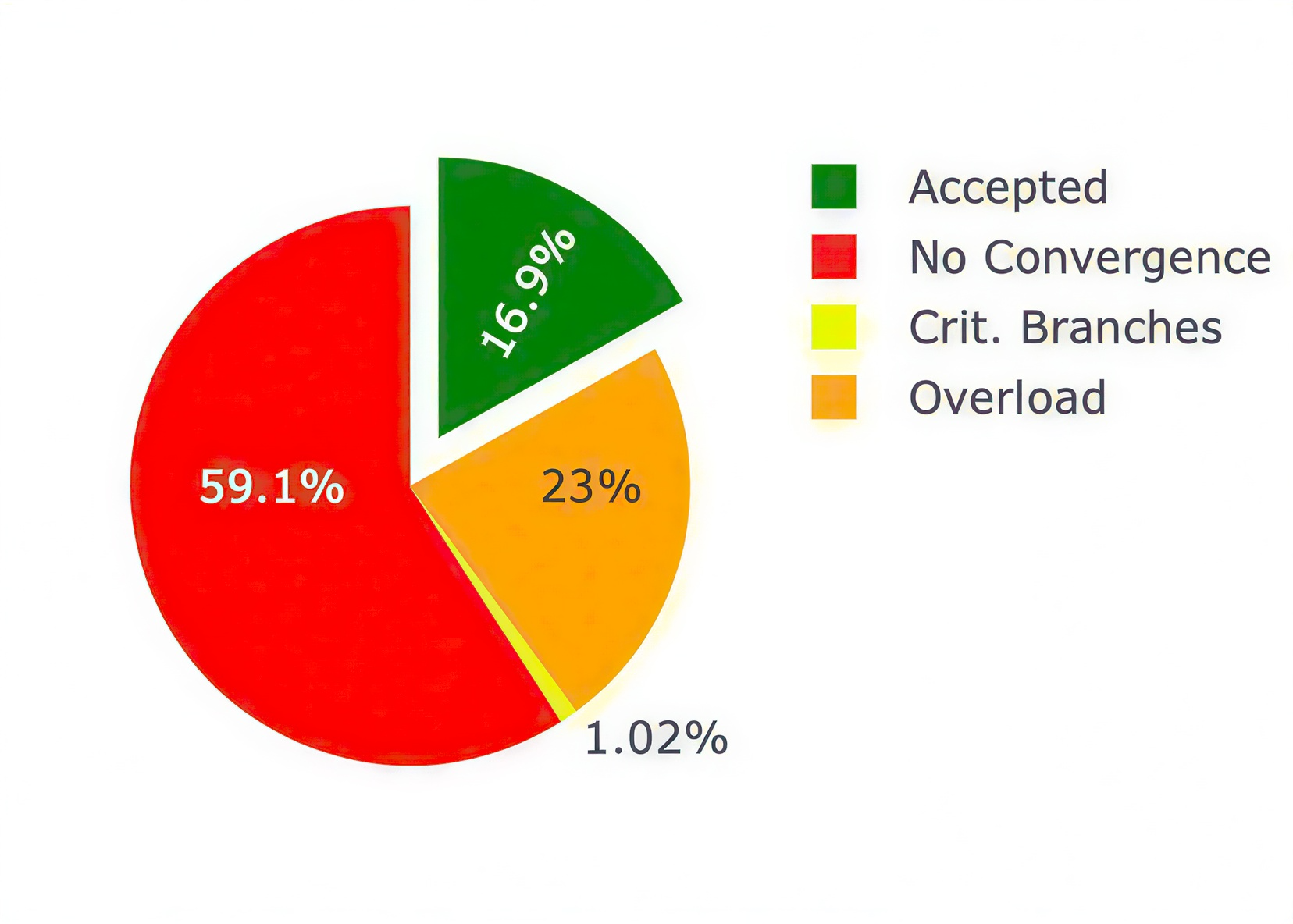}
\caption{Rejection reasons in percent of the DC generated candidates, on the TSO 2 study case}
\label{fig:rejection_reasons}
\end{figure}

To assess whether the heuristics used in the DC stage are effective at producing AC-valid topologies, we furthermore investigate rejection reasons in the AC validation stage. In Figure \ref{fig:rejection_reasons} we plot the percentage of accepted topologies from the subset of all topology candidates that were generated by the DC optimizer. We can see that a relatively small fraction of only 16.9\% topologies passes validation, indicating a mediocre effectiveness of the DC solution. A significant portion of the rejections are due to convergence issues. The reason for why an AC loadflow does not converge can be manyfold. A poor initial guess, voltage collapse, feedback loops in regulating elements, data issues or other problems. We do not further distinguish problems, as it is not trivial to infer which exact cause lies beyond an AC non-convergence \cite{taheri2024ac}. We hypothesize voltage collapse to be a major fraction of the non-convergence cases due to the high influence of topology changes towards this phenomenon. The heavy stress of the day ahead grids further contributes to potential collapse situations, as a branch with high active load is more likely to induce a collapse. 

A quarter of the topologies is rejected due to worsened loadflow results, either because the overload energy did not improve in AC or the number of critical branches increased. These are both criteria that were evaluated in DC and showed an improvement compared to the pre-optimization situation, as otherwise the topology would not be forwarded. Hence, the mismatch in these criteria must be caused by inaccuracies in the DC linearization. 

\subsection{Evolution over time}

\begin{figure*}
    \centering
    \includegraphics[width=1.\linewidth]{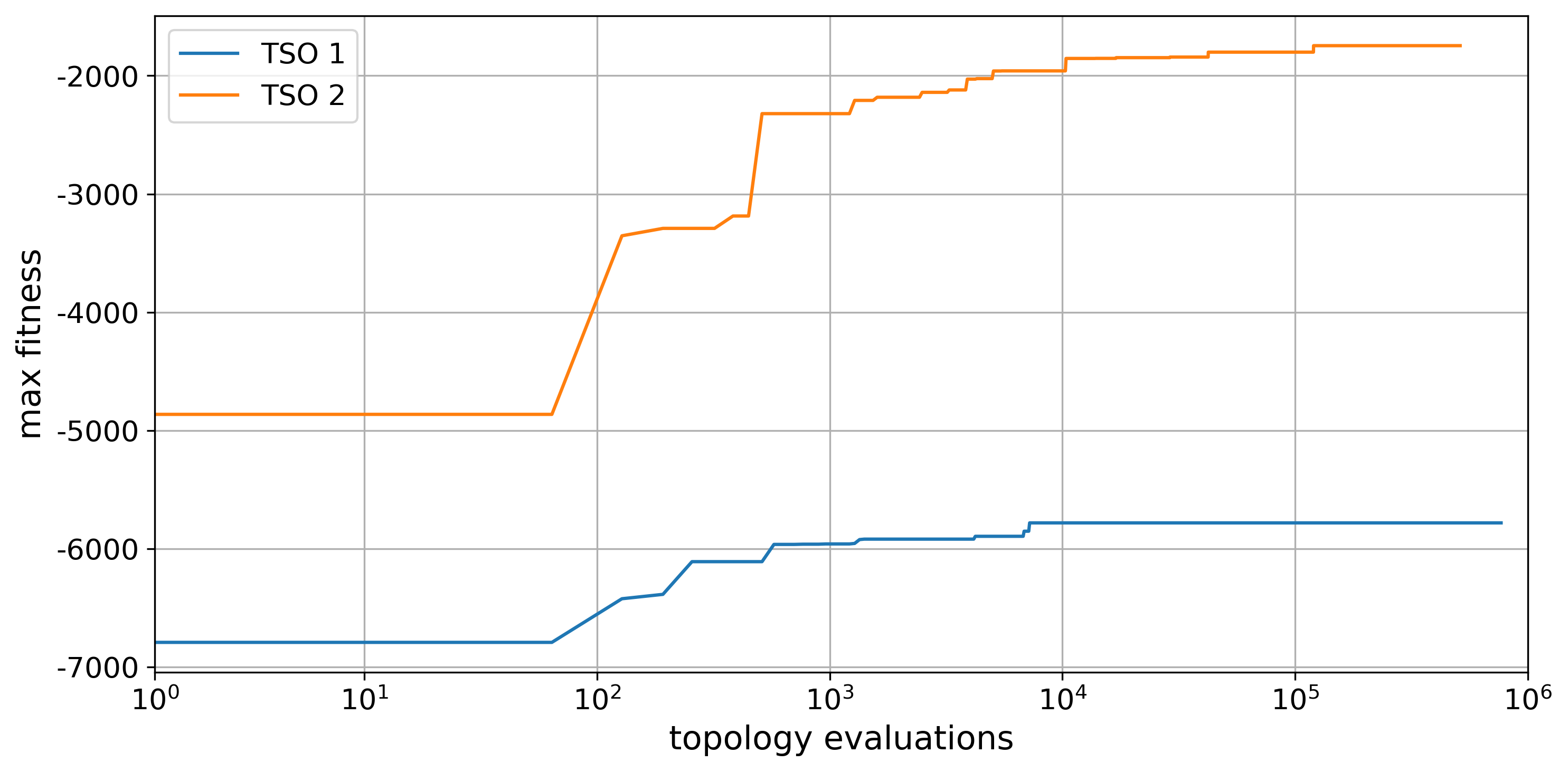}
    \caption{The maximum fitness ($\lambda_{f1}$ for TSO 1 and $\lambda_{f2}$ for TSO 2) in the repertoire after progressively evaluating more candidate topologies. The x axis is displayed in log scale to highlight the improvements at the start of the optimization.}
    \label{fig:fitness_over_time}
\end{figure*}

To assess whether the allocated runtime is enough to find good topologies, we plot the best fitness in the DC repertoire over evaluated topologies in Figure \ref{fig:fitness_over_time}. The total runtime for both scenarios was 3 minutes, and after the initial compile of 13 seconds, time corresponds mostly linear to number of evaluated topologies due to no data-dependent loops in the scoring. Our solution evaluated $768,064$ topologies on the TSO 1 grid and $512,128$ on the TSO 2 grid. The faster runtime on TSO 1 despite the larger grid size is due to the absence of busbar outage computations, which dominate the TSO 2 runtime. To highlight early improvements, we choose a log scale for the x axis. Maximum fitness improves significantly within the first $10,000$ evaluated topologies on both grids. The TSO 1 grid stagnates after this, while the TSO 2 grid still sees improvements up to $180,224$ topologies. This is coherent with the grid situation as outlined in Table \ref{tab:overload_heatmap}, where the TSO 1 grid has many solutions reaching the same terminal fitness. The TSO 2 grid has a more complex tradeoff. After around a fourth of the total runtime, both optimizations stagnate with respect to maximum fitness, however improvements along other parts of the pareto front might still happen. Note that after the initial batch of $64$ topologies, both optimizations did not yet find any improvements over the starting overload. This indicates that the initial seeding of the repertoire with only the pre-optimization grid might be suboptimal.

\section{Conclusion}

In this work, we presented an end-to-end solution for large-scale \ac{TTO} that meets the stringent runtime requirements of operational congestion management. By combining extensive preprocessing, a fully GPU-native DC optimization loop, and a selective AC validation pipeline, we demonstrated that it is feasible to explore an otherwise intractable combinatorial search space within minutes. The use of an accelerated MapElites algorithm enables not only fast convergence but also the illumination of a diverse Pareto front that explicitly exposes the trade-offs between congestion relief and operational risk. The proposed solution enables the first industrialization of full-scale \ac{TTO} known to us at any European \acp{TSO}.

Our results on realistic transmission grid models from two European \acp{TSO} show that thousands of topology candidates per second can be evaluated under the DC approximation. Good results are often found early in the DC search, making the AC validation the main optimization bottleneck of the approach. Despite high rejection rates, the found AC-validated results lead to substantial reductions in N-1 overload energy. However, it is clear that future work is needed to improve the approach.

The importing stage takes up half of the total runtime. However, most of the work performed in that stage could be executed before market clearing offers correct load and generator setpoints, ahead of time. Loading the grid file, preprocessing the action set and generating the PTDF matrix can happen before injections are known. Furthermore, the AC validation could be sped up. Improvements in the parallelization for small instances or a per-topology parallelization to make use of available CPU cores could bring speedups. Furthermore, the potential of GPU-accelerating also the AC stage could be explored \cite{fu2020gpubased}.

% Rejections
The rejection analysis in the AC validation stage highlights the known limitations of DC-based heuristics: AC–DC mismatches and convergence failures remain a significant bottleneck and underline the inherent difficulty of topology optimization in heavily stressed grids. A feedback from the AC back to the DC stage or a better linearization such as DC+ \cite{titz2026voltagesensitive} may alleviate the problem.

Future work could also investigate whether including a redispatch optimization within the approach would lower the number of non-converging loadflows, or even improve result quality. Reduced loading on the grid through redispatch by relieving highly strained lines moves the grid further away from voltage collapse. The heuristics used in this work are designed to approximate redispatch costs. However, a true redispatch run would more accurately represent a grid situation where some congestions dominate redispatch costs and should be the prime focus of a topology optimization.

Overall, our results demonstrate that GPU-accelerated quality-diversity optimization is a practical and promising foundation for deploying \ac{TTO} at scale in real-world transmission system operation. Ongoing evaluation at two European \acp{TSO} will show the potential for redispatch cost savings and operator assistance. By releasing our code open-source, we enable future work to directly base on our approach. 

\bibliographystyle{unsrt}
\bibliography{main}

\end{document}